# ROLE OF CONTEXT IN USABILITY EVALUATIONS: A REVIEW

Munesh Chandra Trivedi[1] and Mohammadi Akheela Khanum[2]

[1]DIT School of Engineering, Greater Noida, India
munesh.trivedi@gmail.com
[2]PAHER University, Udaipur, Rajasthan, India
akheela.khanum@gmail.com

## ABSTRACT

*Usability is often defined as the ability of a system to carry out specific tasks by specific users in a specific context. Usability evaluation involves testing the system for its expected usability. Usability testing is performed in natural environment (field) or artificial environment (laboratory). The result of usability evaluation is affected by the environment in which it is carried out. Previous studies have focused only on the physical environment (lab and field) effect on the results but rarely focused on the effect of social environment (people present during testing). Therefore, this study aims to review how important it is to take context into account during usability evaluation. Context is explored through the theory of behaviour settings, according to which behaviour of individuals is strongly influenced by the physical as well as the social environment in which they function. The result of this review indicates that the physical and social context plays a substantial role in usability evaluations. Further, it also suggests that the usability evaluation model should encompass context as an important component in the framework.*

## 1. INTRODUCTION

Now a days when users buy any gadget, be it a mobile phone, laptop, or an ipad, they first check how easy and understandable the gadget functionality is. Whether they can perform the required operations quickly, are they able to understand the icons on the interface easily without any help. In short, the user is concentrating on the usability of the device. Therefore, usability has become an important parameter today. Usability engineering is a set of activities that ideally take place throughout the lifecycle of the product, with significant activities happening at the early stages before the user interface has been designed [1]. "Usability Engineering" is a science which studies how to understand and systematically address the usability demand of a customer. Thus, usability engineering deals with design of Web sites, computer portals, computer keyboard design, car dashboard design, TV remote key layouts, washing machine front panel layout, etc [2].

### 1.1 Usability

Usability is most often defined as the ease of use and acceptability of a system for a particular class of users carrying out specific tasks in a specific environment. Ease of use affects the user's performance and their satisfaction, while acceptability affects whether the product is used [3]. Nielsen [4] considers that the usability of a system can have five quality components:

(i) Learnability: how easy is it for the users to accomplish basic tasks the first time they encounter the design?
(ii) Efficiency: once users have learned the design, how quickly can they perform tasks?

(iii) Memorability: when users return to the design after a period of not using it, how easily can they reestablish proficiency?

(iv) Errors: how many errors do users make, how severe are these errors, and how easily can they recover from the errors?

(v) Satisfaction: how pleasant is it to use the design?

Usability has been defined by the International Standards Organization (ISO) as "the extent to which the product can be used by specified users to achieve specified goals with effectiveness, efficiency, and satisfaction in a specified context of use" [5].

## 1.2 Usability Evaluation

Usability evaluation is an important part of today's software development process as it can help improve the usability of systems under development. Usability evaluations can save money, time and effort if introduced into the process correctly and at the right time [1]. The basic aim of usability evaluation is to improve the usability of products. Through usability evaluation possible weaknesses with regards to a system's usability with the involvement of actual users can be identified. Usability evaluation involves presenting the users with some tasks which are reflective of the future system use. The results of a usability evaluation can be represented in different forms, such as error rates, time taken to complete the task, and number of usability problems found. Usability evaluation is generally carried out in usability laboratories (in-vitro) and in some cases can be carried out in field (in-situ). Holzinger [6] divided the usability evaluation techniques into inspection methods (without end users) and test methods (with end users). Table 1 depicts this division.

Table 1: Categorization of usability evaluation techniques

Source: [6]

| Inspection Methods | Test Methods |
|---|---|
| **Heuristic Evaluation** | Thinking Aloud |
| **Cognitive Walkthrough** | Field Observation |
| **Action Analysis** | Questionnaires |

Heuristic evaluations are expert evaluations of products or systems, including information systems and documentation. They're conducted by usability specialists, domain experts, or preferably by "double experts" with both usability and domain experience [7]. Advantage of evaluation is that it can produce results in a limited time because it does not involve time-consuming participant recruiting. The disadvantage is that the results of heuristic evaluation cannot be fully trusted as no real users are involved.

Cognitive walkthrough is a task-oriented method by which the analyst explores the system's functionalities; that is, it simulates step-by-step user behavior for a given task. It emphasizes cognitive issues, such as learnability, by analyzing the mental processes required of the users [6]. Advantages include independence from end users and a fully functioning prototype, helping designers to take on a potential user's perspective, effective identification of problems arising from interaction with the system, and the ability to help to define users' goals and assumptions. Disadvantages of cognitive walkthrough include possible tediousness and the danger of an inherent bias due to improper task selection, emphasis on low-level details, and non involvement of the end user [6].

Action analysis involves a walkthrough of the actions a user will perform with regard to physical, cognitive, and perceptual loading. Advantages include precise prediction of how long a task will take, and a deep insight into users' behavior. Disadvantages of action analysis include it is very time-consuming and requires high expertise [6].

With the think aloud protocol, areas where a user is struggling and the reasons for the difficulties are verbally articulated. The usability practitioner uses this information along with other metrics to identify problem areas of the Web site or application being assessed and to devise suggestions for improvement. One of the most common think aloud protocols that usability practitioners engage in today is concurrent think aloud under which the participant is encouraged to "think out loud" while working on a task [8].

Field observation is the simplest of all methods. It involves visiting one or more users in their workplaces. Notes must be taken as unobtrusively as possible to avoid interfering with their work. Observation focuses on major usability catastrophes [6].

Questionnaires are indirect usability measures. They don't study the interface directly; rather collects the user's view about the interface. Questionnaires have to be designed by the experts and should cover all the experiences with the interface. In order to validate the results of the questionnaires large number of users has to be assessed.

Usability evaluations cannot be simply based on the results of application of one or more of the above techniques. Many aspects of context, such as the users, the location, and the culture, all of which can be important during the evaluations, have to be taken into account. Therefore, in what follows we explore what context is? What aspects form the context? And what is the role played by the context in usability evaluations?

## 2. MOTIVATION

In order to understand the relation between the context and the individual's behaviour we chose the widely accepted theory of behaviour settings. The theory of behaviour settings was introduced by Roger Garlock Barker in late 1940s [9]. He continuously collected empirical data from a small town in Kansas with less than 2000 people from 1947 through 1972 based on which he developed the theory of behaviour settings. Behaviour setting theory proposes that there are specific, identifiable units of the environment, the physical and social elements, which are combined into one unit, which have very powerful influences on human behaviour [10]. A behaviour setting is a naturally occurring unit of the environment at the molar (perceived as wholes as opposed to parts) level, recognized by its inhabitants, that is, people perceive that they conduct their lives inside behaviour settings [11;12;9]. Barker [11] observed that psychology is charged with making sense of both the psychological and the ecological environment. He examined that the distinction between human psychology and his ecological environment was difficult. Therefore, Barker focused on molar human behaviour rather than individual units. For example, he interpreted the act of buying a stamp as an entire unified behaviour not broken down into micro acts that followed the stamp buyer through the myriad of smaller components of the total act [9]. The interface of the ecological and molar behaviour creates ecological units [11]. According to Schoggen [9] these units arise simultaneously in physical, social, psychological and behavioural realms and share three common attributes:

1. They are self-generated as opposed to resulting from the observer's or researcher's interest or manipulation.
2. They have a time/space locus.
3. They have a boundary separating the internal pattern of the unit from the external pattern of the surround.

A behaviour setting is a pattern of ecological units and consists of "standing pattern of behaviour" Barker [11]. Barker described the standing pattern as a milieu (settings), circumjacent, and synomorphic or fitting to the behaviour [9]. The behaviour is happening in a milieu and milieu matches the behaviour.

The close interrelation of location/settings and people as seen through the Barker's theory of behavior settings could be an indication that context is important and plays a vital role in influencing the results of usability evaluations.

## 3. CONTEXT IN USABILITY EVALUATION

Context is a term defined differently by different people. For example, Brown et al. [13] define context as location, identities of the people around the user, the time of the day, season, and temperature. Ryan et al [14] define context as the user's location, environment, identity and time. Hull et al [15] included the entire environment by defining context to be aspects of the current situation. Schilit et al [16] claim that the important aspects of the context are: where you are, who you are with, and what resources are nearby. Dey et al [17] define context to be the user's physical, social, emotional or informational state.

Taking into consideration the various definitions of context, we found that one of the likely definitions of the context would be *"context is anything which has an effect on the human behaviour"*. When evaluating the usability of any system, the behaviour of the user is very important. The factors which may affect the user behaviour needs to carefully considered because the result of usability evaluations may vary in different settings where the user may exhibit varying behaviours.

Product usability doesn't take place in a vacuum; rather, it happens in context [8]. It is not meaningful to talk simply about the usability of a product, as usability is a function of the context in which the product is used. The characteristics of the context (the users, tasks, and environment) may be as important in determining usability as the characteristics of the product itself. Changing any relevant aspect of the context of use may change the usability of the product [19]. To understand the role of context in usability evaluation, it is necessary to examine what context is, and what aspects it comprise of. Many studies report that just physical location is the context. Context can be the cultural context [20], organizational context, technological context or social context [21]. Since our focus is on the physical and social context, we explore each one of them in the next two sections.

### 3.1 Physical Context

Physical context comprise of the physical surroundings of the users, it is the location, the place where usability evaluation takes place. Physical context usually refers to the environment in which user is tested. Natural environment is the location of actual use of the system being tested. Usability evaluation taking place in natural environment is called as the field testing. Artificial environment is the simulation of natural environment, sometimes referred to as the controlled environment. Usability evaluation carried out in artificial environment is also referred to as laboratory testing.

Traditionally laboratory experiments are employed to evaluate the usability of computer systems, and to improve the understanding of usability. Laboratory testing takes place in a controlled environment with the experimenter in control of assignments of subjects, treatment variables and manipulation of variables. It is possible to employ facilities for collection of high-quality data such as video recording of the display and user interaction [22]. Razak et al. [23] states that the evaluation done in the laboratory has several advantages. First, the conditions for conducting research can be controlled. Secondly, all the participants experience same setting leading to higher quality data. Laboratory studies allow the researchers to focus on specific phenomena of interest and facilitate good data collection. Laboratory testing has received both appreciation as well as criticism. The laboratory evaluations do not simulate the context when usability testing is done with mobile phones, because laboratory settings lack the desired ecological validity [24]. Even though the adequateness of laboratory evaluations is questioned,

71% of the mobile device evaluations were done in laboratory settings [25]. Similar claim also comes from Park & Lim [26] where they state that simulating the use settings is very hard, time consuming, expensive and lacks contextual factors.

Field testing takes place in a more natural setting. An artificial setting supports control but lacks realism whereas a natural setting supplies realism but makes control more difficult [27]. Oh & Kim [28] claims that to test children requirements and issues with the everyday use of technology, field evaluation would be a better choice. This claim is also supported by Xu et al. [28] in search for evaluation methods for children's tangible technology, where they found that location plays a large part in how children behave, children felt more at ease and focused when they are at school.

### 3.2 Social Context

A major part of the context of a usability evaluation is the people involved. This is also often referred to as the social context of the usability evaluation. People involved in usability evaluations can be the evaluators, the test monitors, the users, and other people who may not be directly involved with the evaluation, however, their presence can have a substantial effect on the results of usability evaluations. Stoica et al. [30] found that while laboratory evaluations give excellent data, the context and the surroundings as well as other people around also play an important role. One of the purposes of creating a social context in usability evaluations is to facilitate effective and efficient evaluations. Creating a proper social context can potentially diminish some of these challenges [31]. Although social context is considered important, only little research has been done to identify how it influences usability evaluations.

## 4. RELATED WORK

Research in usability evaluation is very old and still is an active area of research. Literature provides the evidence that many studies till date have focussed on the various usability evaluation issues and challenges. Studies addressing the contextual issues during usability evaluations can be found in the literature. We reviewed some of the previous work specifically concentrating on the physical and social context in usability evaluations. The related work is classified based on the context they target.

### 4.1 Based on physical context

The importance of physical context in usability evaluations have been researched for a long. Out of the many factors that can effect usability evaluations, physical context is considered to directly influence the behaviour of the people involved in the usability evaluations. The physical context may include the location, the temperature, the time, the light etc.

Tsiaousis & Giaglis [32] examined the effects of environmental distractions on mobile website usability. They proposed a model hypothesizing on the effects of environmental distractions on the usability of mobile sites. They categorized the environmental distractions into auditory, visual and social. A preliminary test on 20 users was conducted to investigate the effect of environmental distractions on mobile website usability. Results confirmed that environmental distractions have direct effect on mobile website usability.

Hummel et al. [33] developed a mobile context-framework based on a small wireless sensor network, to monitor environmental conditions such as light, acceleration, sound, temperature, and humidity during the usability experiments. User experiments have been conducted in a laboratory with seven test persons where the environmental conditions were changed. Under varying environmental conditions the performance of the users on the average was decreased in terms of higher error rates and delays.

Kaikkonen et al. [34] carried out usability testing of mobile consumer application in two environments: in a laboratory and in a field with a total of 40 test users. Results indicate that conducting a time-consuming field test may not be worthwhile when searching user interface flaws to improve user interaction. They found that field testing is worthwhile when combining usability tests with a field pilot or contextual study where user behaviour is investigated in a natural context.

Razak et al. [23] conducted usability testing with children in both laboratory and field. Drawing applications were tested in their preschool and an educational game was tested in the usability laboratory. The results indicate that field study is more suitable for understanding children experience with technology than it is with testing for usability problems and laboratory study is more suitable for evaluating user interfaces and interaction with the application than it is with understanding children's experience.

Andrrzejczak & Liu [35] examined the effect of location on the user's stress level during usability evaluation. User stress levels were assessed by Spielberger's State-Trait Anxiety Inventory; using the paper survey's baseline and experimental stress scores. In addition, user performance data was recorded through task times and subjective user assessments. The data suggested no significant differences exist between participant data in both baseline and experimental anxiety scores. This implies that remote testing as a cost-efficient way to conduct user testing, may be a viable alternative to traditional lab testing without altering the test's effectiveness.

Madathil [36] performed a synchronous remote usability test using a three-dimensional virtual world, and empirically compared it with WebEx, a web-based two-dimensional screen sharing and conferencing tool, and the traditional lab method. The results suggest that virtual lab method is as effective as the traditional lab and WebEx based methods in terms of the time taken by the test participants to complete the tasks and the number of higher severity defects identified. Test participants and facilitators alike experienced lower overall workload in the traditional lab environment than in either of the remote testing environments.

Baillie & Schatz [37] evaluated a multimodal mobile application through a combination of laboratory and field studies. The users were given a set of four action scenarios to be performed. The results were surprising; only one action scenario was completed in the time frame whereas three out of four action scenarios were completed in lesser time. Error rates were higher in lab than in the field. The reason for such performances by the users could be that the users feel more relaxed in the field.

## 4.2 Based on the social context

The effect of social context in usability evaluations can be understood by examining some of the previous works in the domain. Work by Benedikte et al. [38] involved conducting usability testing with 60 children with three setups where children apply think-aloud or constructive interaction in acquainted and non-acquainted pairs. The results show that the pairing of children had impact on how the children verbalized and collaborated in pairs during the testing sessions. The children in pairs had a high level of verbalization, but often they were more talking aloud than actually thinking aloud. The acquainted dyads were significantly more satisfied with their own performance and they did not feel it demanded a lot of effort for them.

Study by Jacobsen et al. [39] examines the evaluator effect in the usability tests. In their study four HCI research evaluators, all familiar with the theory and practice of usability analyzed four video tapes. The evaluators were asked to detect and describe all problems in the interface based on analyzing the four tapes in a preset order, without any time constraints. The results indicate that only 20% of the 93 unique problems were detected by only a single evaluator. Severe problems were detected by more often by all four evaluators (41%) and less often by only one evaluator (22%), however, the evaluator effect remained substantial.

Evaluator effect has also been probed by Hertzum and Jacobsen [40], where they found that different evaluators evaluating the same system with same usability evaluation methods detect substantially different sets of usability problems in the system.

van den Haak & de Jong [41] analyzed the interaction between test monitor and participants in concurrent think aloud (CTA) method and constructive interaction (CI) test. The results indicate that the presence of the test monitor has most notably affected the CTA participants but has also has its impact on CI participants. They found that a more serious threat to the validity of both the CTA method and CI method was that the participants acknowledged the test monitor as an evaluator of their actions.

## 5. DISCUSSION

We reviewed many works which took into consideration the importance of the physical as well as the social contexts in usability evaluations. The summary of the relevant work is given in table 2. The entries in the table depict the type of context that was addressed in the work, whether context affected the outcomes of the work and the target participants. Physical context in the table refers to both the laboratory evaluation and field evaluation. Social context refers to the people involved in the test. Both refer to the physical and the social context taken together.

Table 2: Summary of the relevant work

| Work | Context used | | | Does context effects the results | | | Test participants | | Application tested |
|---|---|---|---|---|---|---|---|---|---|
| | Physical | Social | Both | Yes | No | Sometimes | Children | Adults | |
| [32] | ✓ | ✗ | ✗ | ✓ | ✗ | ✗ | ✗ | ✓ | Mobile website |
| [33] | ✓ | ✗ | ✗ | ✓ | ✗ | ✗ | ✗ | ✓ | Mobile |
| [34] | ✓ | ✗ | ✗ | ✓ | ✗ | ✗ | ✗ | ✓ | Mobile commerce application |
| [23] | ✓ | ✗ | ✗ | ✓ | ✗ | ✗ | ✓ | ✗ | Drawing application |
| [35] | ✓ | ✗ | ✗ | ✗ | ✓ | ✗ | ✗ | ✓ | Commercial website |
| [36] | ✓ | ✗ | ✗ | ✗ | ✗ | ✓ | ✗ | ✓ | E-commerce website |
| [37] | ✓ | ✗ | ✗ | ✗ | ✓ | ✗ | ✗ | ✓ | Multimodal mobile application |
| [38] | ✗ | ✓ | ✗ | ✓ | ✗ | ✗ | ✓ | ✗ | Mobile phone |
| [39] | ✗ | ✓ | ✗ | ✓ | ✗ | ✗ | ✗ | ✓ | Video tapes |
| [41] | ✗ | ✓ | ✗ | ✓ | ✗ | ✗ | ✗ | ✓ | Online library catalogues |

Physical context is addressed by many studies whereas, the social context in usability evaluations is explored by a fewer studies. Most of the studies show that the context (physical and social) has an impact on the outcomes of the evaluations. The focus on physical context in usability evaluations indicates that this aspect of context is considered important. The choice is between evaluating in an artificial setting such as the laboratory or in a more natural setting through field evaluations. Each has its own strengths and weaknesses.

Jones & Marsden [42] state that the social context in usability evaluation is equally important. The main objective of creating a social context in usability evaluation is to facilitate effective and efficient evaluations. Pardo et al. [43] examined the effect of teacher's involvement in usability testing with children. They found that children cannot provide proper feedback on the learning goals they have not experienced before. In such cases the input of the other stakeholders such as the teachers will be beneficial. Some studies also reported that, compared to evaluations involving adult participants, studies involving children are mostly affected by the context. Children show varying behaviour when they are tested in the laboratory environment and when they are tested in the field. They feel relaxed and confident in their own environment. Children also show varying behaviour when they are accompanied by the people they are well acquainted with and with the people they are not acquainted with. Almost 80% of the work that we surveyed shows a clear impact of the context on the outcomes of the usability evaluations. Studies concerning the mobile applications usability testing are the most affected by the context. With mobile applications, better insight to the usability problems can be unveiled in the field than in the lab. None the less, usability evaluations involving adults may also be effected by the context.

While previous research studies in usability evaluations have largely focused on the physical aspects of context in usability evaluations, work on social aspects of context is scarce. Behaviour of the test participants is affected by the context which may in turn affect the results of usability evaluations. Therefore, behaviour and settings are inseparable units as claimed by Barker's theory of behaviour settings. More research taking into account both the social and physical context together is needed to uncover the importance of context in usability evaluations.

## 6. CONCLUSION

This paper is a review of the role of context in usability evaluations. Essentially the role of the physical and social context in the literature was examined. Roger Barker's theory of behaviour settings can be a strong motivation for investigating context. Behaviour setting theory considers behaviour and the settings are inseparable units .Behaviour setting theory has been widely cited for its potential applications to community psychology. While the theory has broad applicability and a strong empirical base, research on it is limited. Future research will focus on examining the elements of physical and social context in an attempt to understand the influence of context in usability evaluations. However, there is a lack of coherence in understanding the context through psychological perspective. The research on context is scattered and scarce, lacking a unifying overview. Therefore, an understanding of the influence of context and how it impacts the process of usability evaluation is needed. This study has formed the basis for our future work in exploring the role of context in usability evaluations. It has also established the line of investigation that is needed to move forward in developing a usability evaluation framework encompassing context.